\documentclass[runningheads]{llncs}

\usepackage[T1]{fontenc}
\usepackage{graphicx}
\usepackage{xcolor}
\usepackage{hyperref}

\begin{document}

\title{PolyPresentation: A Multimodal AI Platform for Slide-Aware Iterative Presentation Practice}

\titlerunning{PolyPresentation}

\author{
Chen Chen\textsuperscript{1*} \and
Jihao Li\textsuperscript{2*} \and
Zhiyuan Wen\textsuperscript{1,2(\textdagger)} \and
Tianhui Zhang\textsuperscript{3} \and
Di Zou\textsuperscript{4} \and
Jiannong Cao\textsuperscript{1,2}
}

\authorrunning{C. Chen et al.}

\institute{
The Hong Kong Polytechnic University, Hong Kong, China \\
\email{\{chen03.chen, tianhui.zhang\}@connect.polyu.hk, \{jihao.li, zhiyuan.wen, daisy.zou, jiannong.cao\}@polyu.edu.hk}
}

\begingroup
\renewcommand{\thefootnote}{}
\footnotetext{%
\scriptsize
\textsuperscript{*} Equal contribution.\\
\textsuperscript{\textdagger} Corresponding author.\\
\textsuperscript{1} Department of Computing;
\textsuperscript{2} Institute for Higher Education Research and Development;
\textsuperscript{3} Department of Data Science and Artificial Intelligence;
\textsuperscript{4} Department of English and Communication.
}
\endgroup

\maketitle

\begin{abstract}
Presentations are essential for students, researchers, and professionals to communicate ideas persuasively, yet delivering them effectively requires repeated practice that coordinates content, delivery, visual materials, and audience interaction. Existing AI-assisted rehearsal tools provide scalable feedback, but they often treat presentations as single-run delivery performances, offering limited support for linking feedback to the slide deck or planning what to practice in the next iteration. To address this gap, we introduce PolyPresentation, a multimodal AI platform for slide-aware iterative presentation practice. PolyPresentation organizes slide-by-slide practice, full rehearsal, audience Q\&A, and feedback into a unified practice loop, using slide-grounded evidence to help presenters diagnose performance issues and prepare for subsequent practice. We evaluate PolyPresentation through a rubric-based comparison with four baseline systems on 20 academic presentation rehearsals, and additionally assess its alignment with human ratings. Results suggest that PolyPresentation provides more actionable, context-aware, and practice-oriented support for improving presentations. The demonstration video is available at \url{https://youtu.be/MmWj9O_PJxw}.

\end{abstract}

\section{Introduction}

Presentations are essential for students, researchers, and professionals to communicate ideas persuasively, but delivering them effectively requires repeated practice. Presenters must coordinate content, delivery, visual materials, timing, and audience interaction across the talk. In practice, they often rely on self-review or human coaching. Self-review is accessible but subjective and incomplete, while human coaching can provide richer feedback but is limited by cost, availability, and timing. These limitations motivate scalable practice support that can help presenters understand their performance and prepare for subsequent practice.

Recent work has advanced computational support for presentation and public-speaking practice. Speech and multimodal coaching systems can detect delivery behaviors such as pace, pauses, posture, gesture, and timing \cite{tanveer2015rhema,schneider2015presentation,ochoa2024openopaf}; virtual-audience and interview-practice systems make rehearsal more socially situated \cite{chollet2015virtual,leong2024highstakes}; and slide-generation and evaluation research has improved how presentation materials are created and assessed \cite{sun2021d2s,fu2022doc2ppt,zheng2025pptagent,chen2026presentbench}. Recent LLM-based coaching tools further enable flexible feedback \cite{chen2025presentcoach}. However, many tools still treat presentations as single-run delivery performances, with feedback only weakly connected to the slide deck and limited guidance for what to practice next. Yet effective presentation practice is inherently slide-aware and iterative, requiring presenters to coordinate delivery, content coverage, visual support, and audience response across repeated practice.

To address this gap, we introduce \textbf{PolyPresentation}, a multimodal AI platform for slide-aware iterative presentation practice. PolyPresentation organizes practice, rehearsal, simulated audience Q\&A, and feedback into a unified practice loop. Rather than evaluating a full rehearsal only as a completed delivery performance, the platform grounds feedback in the evolving slide context, links feedback to supporting evidence, and helps presenters decide what to improve before the next practice attempt. PolyPresentation is therefore designed as formative support for iterative improvement rather than as a one-time scoring tool.

We evaluate PolyPresentation on 20 academic-conference presentation rehearsals through human-reference alignment and feedback-quality comparison. For human alignment, our platform shows strong agreement with human ratings. In comparison with four baseline systems, PolyPresentation achieves the highest overall feedback-quality score, with particular strengths in Coverage, Actionability and Organization. These results suggest that PolyPresentation produces assessments broadly consistent with human judgment while translating rehearsal evidence into actionable guidance for subsequent practice.

To summarize, this paper makes the following contributions:
\begin{enumerate}
    \item We present PolyPresentation, a multimodal AI platform that integrates practice, rehearsal, simulated audience Q\&A, and feedback into a slide-aware iterative presentation practice workflow.
    \item PolyPresentation introduces a slide-grounded evidence construction approach that links presentation feedback to both deck context and rehearsal activity.
    \item PolyPresentation is evaluated through baseline comparisons and alignment analysis against human ratings, examining both feedback quality and rubric-assessment consistency.
\end{enumerate}

\section{Related Work}

\subsection{Presentation Coaching and Multimodal Feedback}

Automated presentation coaching systems aim to make rehearsal more frequent, private, and measurable. Real-time systems such as Rhema and Presentation Trainer provide feedback on public-speaking behaviors during practice, including pace, volume, posture, gesture, and timing \cite{tanveer2015rhema,schneider2015presentation}. More recent multimodal learning-analytics work has estimated presentation competence from audiovisual cues, studied transfer across assessment settings, and improved reproducibility through open oral-presentation feedback platforms \cite{sumer2021estimating,su2023transfer,ochoa2024openopaf}. Virtual-audience and interview-practice systems further make rehearsal more
socially situated through audience simulation, adaptive dialogue, and multimodal assessment \cite{chollet2015virtual,leong2024highstakes,wen2026polyinterview}.

These systems show that computational feedback can help presenters notice delivery problems during rehearsal. However, such feedback often treats pace, pauses, posture, or gaze as separate indicators. Without slide context, presenters may struggle to decide whether an issue should be addressed through delivery practice, content coverage, slide revision, or audience-preparation strategies. This motivates slide-aware support that connects delivery evidence with deck context and next-round practice.

\subsection{LLM/VLM Feedback and Slide-Aware Presentation Support}

Recent LLM- and VLM-based systems enable more contextualized feedback than numerical dashboards alone. LLMs have been explored for formative feedback and personalized support, while also raising concerns about reliability, over-trust, and weak grounding \cite{kasneci2023chatgpt}. In presentation learning, PresentCoach uses a dual-agent design to generate exemplars and provide interactive rehearsal feedback \cite{chen2025presentcoach}. VLMs further allow systems to reason over slide screenshots and selected video frames, both of which are central to presentation practice.

A related stream studies iterative content authoring and presentation-material
generation. InteractiveSurvey supports user refinement of intermediate structures and generated content throughout an LLM-assisted authoring process \cite{wen2025interactivesurvey}, while D2S and DOC2PPT formulate document-to-slide generation as content selection, summarization, and layout generation \cite{sun2021d2s,fu2022doc2ppt}. Recent agentic systems such as PPTAgent, PreGenie, and SlideGen emphasize editable slide construction,
multimodal review, visual-loop collaboration, and rubric-based evaluation \cite{zheng2025pptagent,xu2025pregenie,liang2025slidegen}; PresentAgent further connects slides with synchronized narration \cite{shi2025presentagent}. This work improves presentation artifacts, but slide evaluation and rehearsal evaluation are still often handled separately. As a result, existing systems do not fully support a slide-aware practice loop that grounds feedback in
rehearsal evidence and helps presenters decide what to practice or revise next.

\begin{figure}[t]
  \centering
  \IfFileExists{workflow-temp.pdf}{%
    \includegraphics[width=\linewidth]{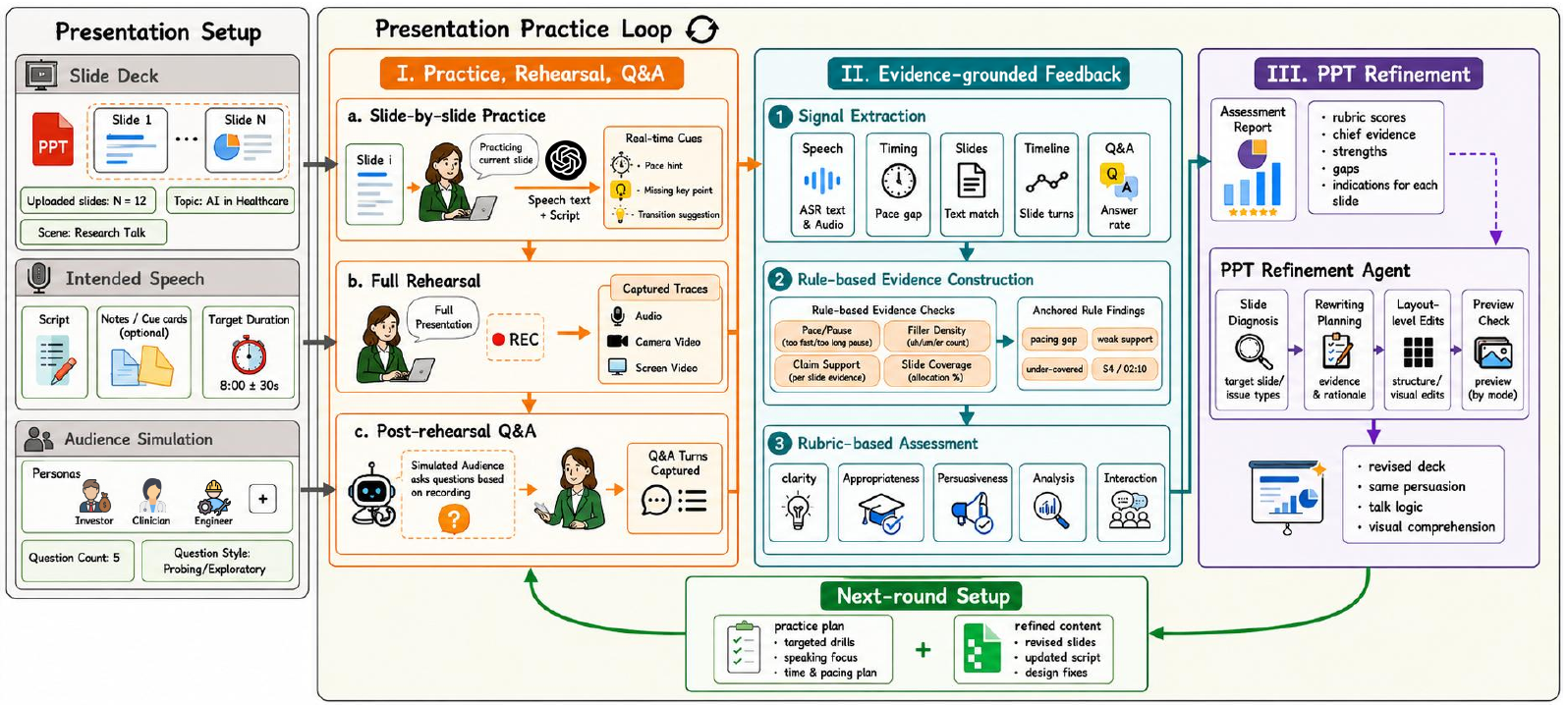}%
  }{%
    \fbox{\parbox[c][5cm][c]{0.95\linewidth}{\centering\textit{[Placeholder:\\polypresentation\_workflow-temp.pdf]}}}%
  }
  \caption{Workflow of PolyPresentation. The platform connects presentation setup, practice, rehearsal, Q\&A, evidence-grounded feedback, and slide deck refinement into a slide-aware iterative practice loop.}
  \label{fig:polypresentation-workflow}
\end{figure}

\section{PolyPresentation}

\subsection{Overview and Workflow}

PolyPresentation is a multimodal AI platform for slide-aware iterative presentation practice. As shown in Fig.~\ref{fig:polypresentation-workflow}, the workflow begins with presentation setup and then proceeds through a three-stage practice loop: (1) practice, rehearsal, and Q\&A, (2) evidence-grounded feedback, and (3) slide deck refinement. During setup, the presenter uploads a slide deck and specifies the presentation context, including the topic, audience, target duration, intended speech or notes, practice goals, and audience simulation settings. PolyPresentation parses the deck into slide-level context, such as titles, bullet text, speaker notes, keyword candidates, and basic text statistics, so later feedback can be interpreted against the actual deck being practiced.

The workflow supports improvement across practice rounds rather than producing a one-shot score. PolyPresentation aligns delivery, timing, slide progression, and interaction records along the slide timeline to construct slide-grounded evidence. This evidence is used to generate feedback, identify next-round practice actions, and guide slide deck refinement when deck-level changes can better support the presenter.

\subsection{Practice, Rehearsal, and Q\&A}

The first stage helps presenters move from local slide preparation to a complete presentation run. During slide-by-slide practice, PolyPresentation provides real-time LLM-based hints (GPT-5) while the presenter is speaking. It uses automatic speech recognition (ASR) to continuously transcribe the presenter's speech, then generates hints based on the current slide content and the elapsed speaking time on that slide. These hints remind presenters of key points to cover, flag possible pacing or transition issues, and suggest concrete next actions. Rather than rewriting the presentation for the presenter, this stage helps presenters notice slide-specific risks and adjust their delivery before recording a full take.

During full rehearsal, PolyPresentation preserves a low-interruption delivery experience while recording the evidence needed for later analysis. Built on LiveTalking\footnote{\url{https://github.com/lipku/LiveTalking}}, the system includes digital avatars during full rehearsal, providing a lightweight sense of audience presence and making the practice experience more engaging for presenters. Meanwhile, PolyPresentation captures the transcript, slide-switch timeline, timing patterns, and available audio, screen, or visual traces. Because each event is timestamped, the platform can later locate what the presenter said, which slide was visible, how long the presenter stayed on the slide, and where pacing or coverage problems emerged.

The post-rehearsal Q\&A component extends practice beyond one-way delivery. Based on the deck, presentation goal, audience profile, and rehearsal evidence, PolyPresentation generates slide-anchored questions that probe clarification, evidence, trade-offs, or implications. Presenter answers are stored as additional evidence for assessing audience readiness, content understanding, and the ability to respond beyond the prepared script.

\subsection{Evidence-grounded Feedback}

The second stage converts practice records into evidence-grounded feedback. PolyPresentation first extracts signals such as transcript segments, slide transitions, timing features, keyword coverage, and Q\&A turns. It then aligns these signals with the slide timeline, so each feedback item can be tied to the slide context in which the issue occurred. This allows the system to reason about problems such as rushed transitions, underdeveloped explanations, weak slide-speech alignment, uneven pacing, or unanswered audience concerns.

PolyPresentation evaluates the structured evidence with a rubric for academic presentation practice. The feedback is formative rather than high-stakes: each score or comment is linked to supporting evidence such as a slide number, timestamped transcript span, timing pattern, or Q\&A turn. To reduce unsupported inference, the system marks unavailable or unreliable modalities as not assessed instead of guessing from incomplete evidence. The final report summarizes strengths, gaps, priority slide moments, a revised-script suggestion, and a rehearsal plan of next-round practice actions, helping presenters understand what to improve and how to continue practicing.

\subsection{Slide Deck Refinement}

The third stage translates selected feedback into deck-level refinements for the next practice round. Using the slide-grounded evidence record, PolyPresentation identifies target slides and determines whether each issue is better addressed through delivery practice, slide revision, or both. For example, a late-transition issue may call for clearer grouping or a presenter-note cue, while low coverage of a key term may require a more explicit slide takeaway or speaking reminder. The refinement agent then proposes revisions under constraints that preserve the presenter's topic, claims, evidence, audience purpose, and overall argument. These revisions may include clearer claim-based titles, simplified bullet structures, improved visual hierarchy, signpost cues, or speaker notes for pacing and Q\&A preparation.

\section{Demonstrations}

\begin{figure}
  \centering
  \begin{minipage}[t]{0.45\textwidth}
    \centering
    \IfFileExists{rehearsal_interface.pdf}{%
      \includegraphics[width=\linewidth,height=3cm,keepaspectratio]{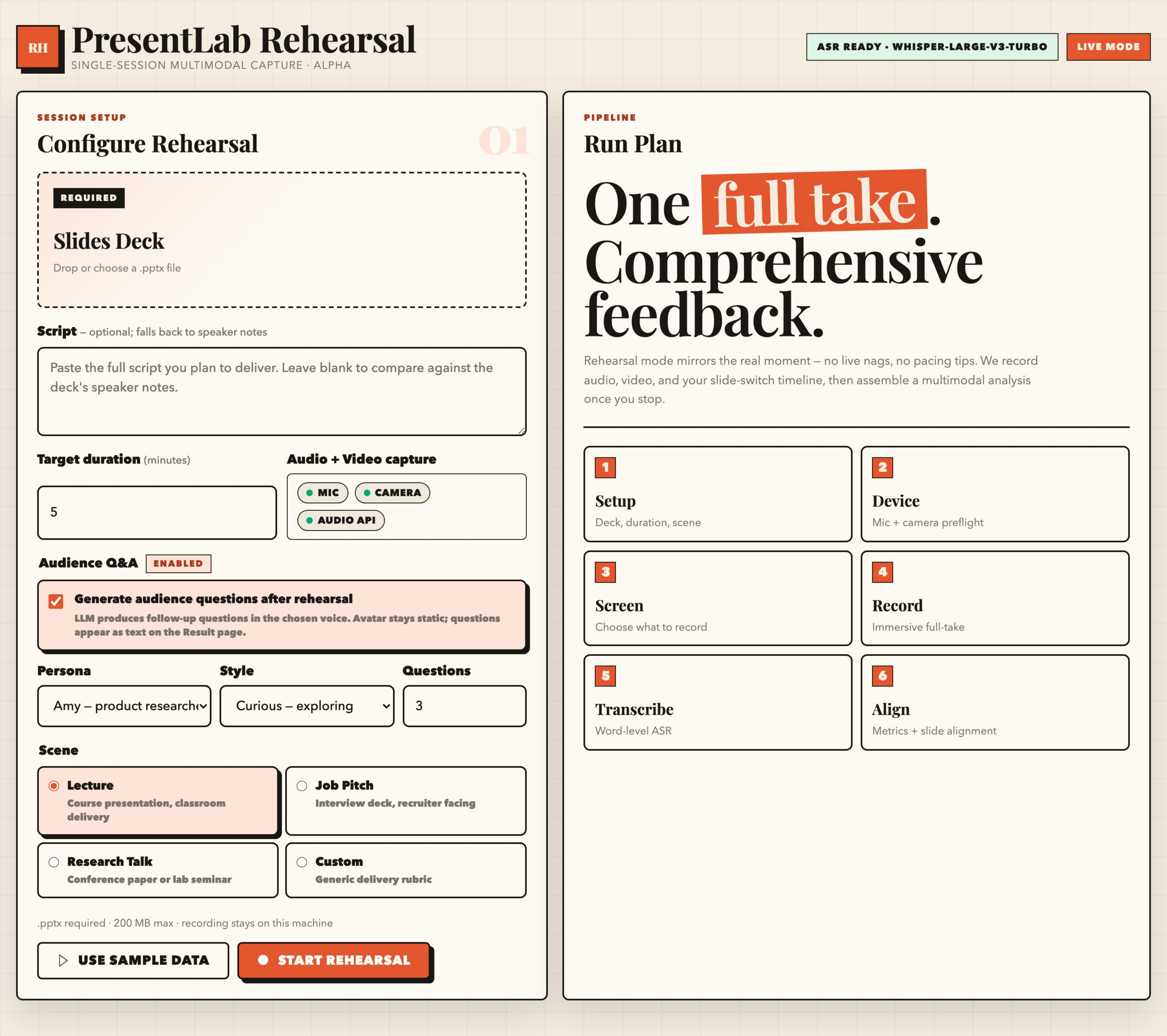}%
    }{%
      \fbox{\parbox[c][3cm][c]{0.95\linewidth}{\centering\textit{[Placeholder:\\rehearsal\_interface.pdf]}}}%
    }
    \parbox{\linewidth}{\centering(a) Rehearsal Mode interface for multimodal capture and live verification.}
  \end{minipage}\hfill
  \begin{minipage}[t]{0.45\textwidth}
    \centering
    \IfFileExists{feedback_interface.pdf}{%
      \includegraphics[width=\linewidth,height=3cm,keepaspectratio]{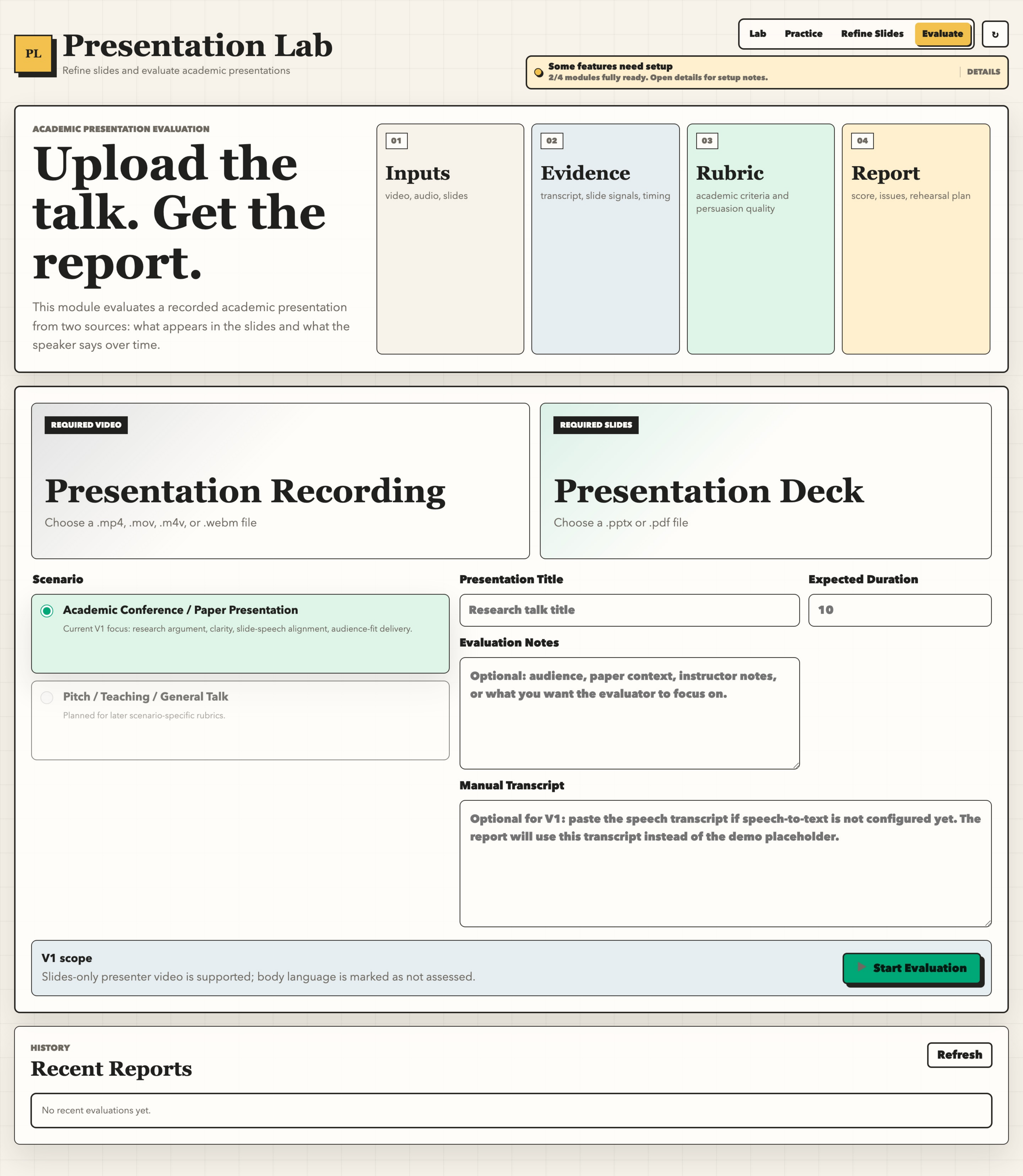}%
    }{%
      \fbox{\parbox[c][3cm][c]{0.95\linewidth}{\centering\textit{[Placeholder:\\feedback\_interface.pdf]}}}%
    }
    \parbox{\linewidth}{\centering(b) Evaluation Dashboard input panel.}
  \end{minipage}
  \caption{Main user-facing interfaces of PolyPresentation.}
  \label{fig:polypresentation-interface}
\end{figure}

Fig.~\ref{fig:polypresentation-interface} shows the two main interfaces. The \textbf{Rehearsal Mode} interface records audio, optional camera, screen capture, and slide-switch events while the presenter runs a low-interruption rehearsal. The \textbf{Evaluation Dashboard} pairs each rubric judgment with its supporting evidence (transcript spans, the slide on screen, rule-derived signals, optional prosody and visual-behavior traces); modalities that are unavailable or unreliable are reported as \textit{not assessed} rather than imputed.

\begin{figure}
  \centering
  \begin{minipage}[t]{0.32\textwidth}
    \centering
    \IfFileExists{demo1.pdf}{%
      \includegraphics[width=\linewidth,height=2.5cm,keepaspectratio]{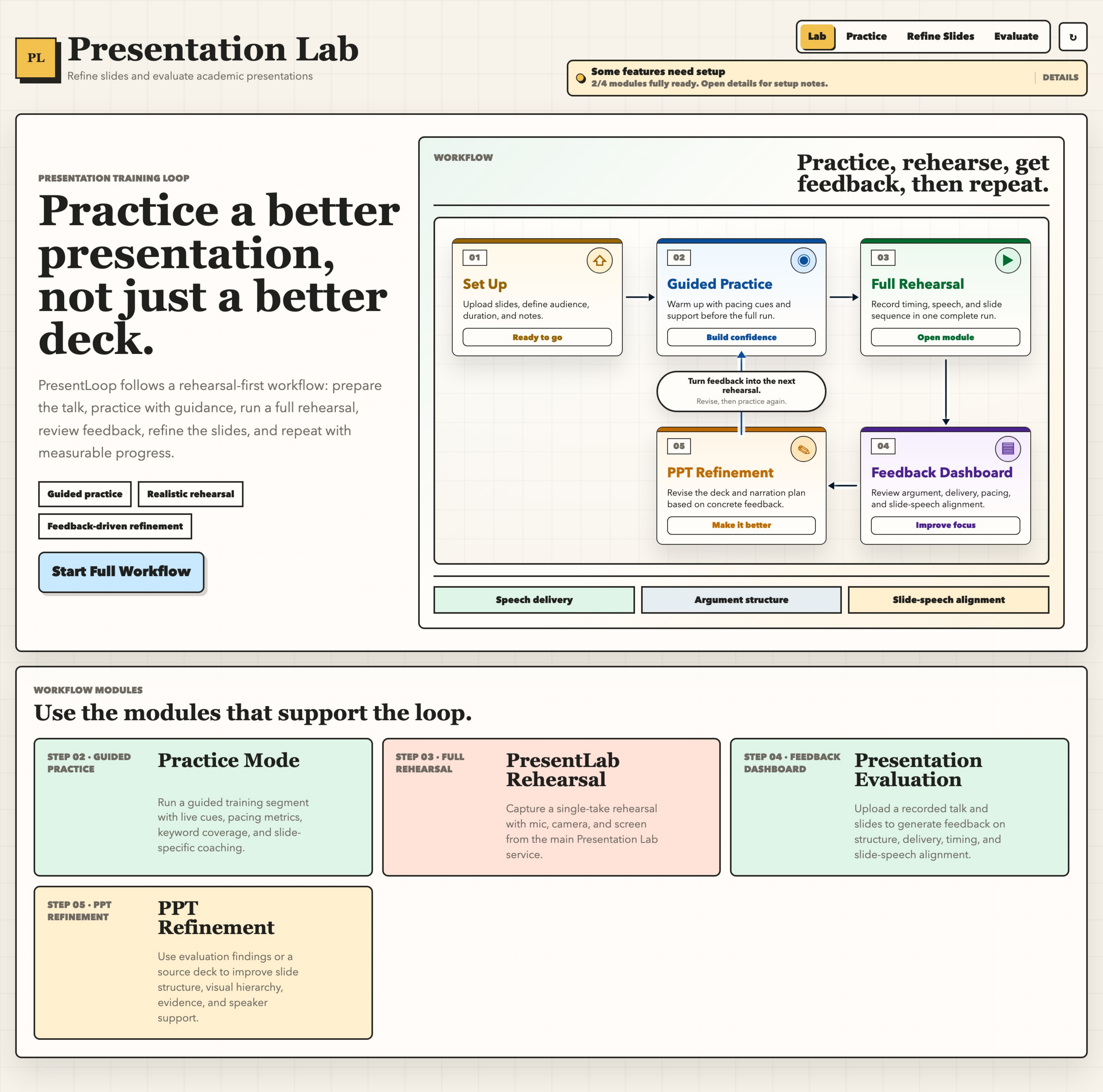}%
    }{%
      \fbox{\parbox[c][2.5cm][c]{0.95\linewidth}{\centering\textit{[Placeholder:\\demo1.pdf]}}}%
    }
    \parbox{\linewidth}{\centering(a) Set Up.}
  \end{minipage}\hfill
  \begin{minipage}[t]{0.32\textwidth}
    \centering
    \IfFileExists{demo2.pdf}{%
      \includegraphics[width=\linewidth,height=2.5cm,keepaspectratio]{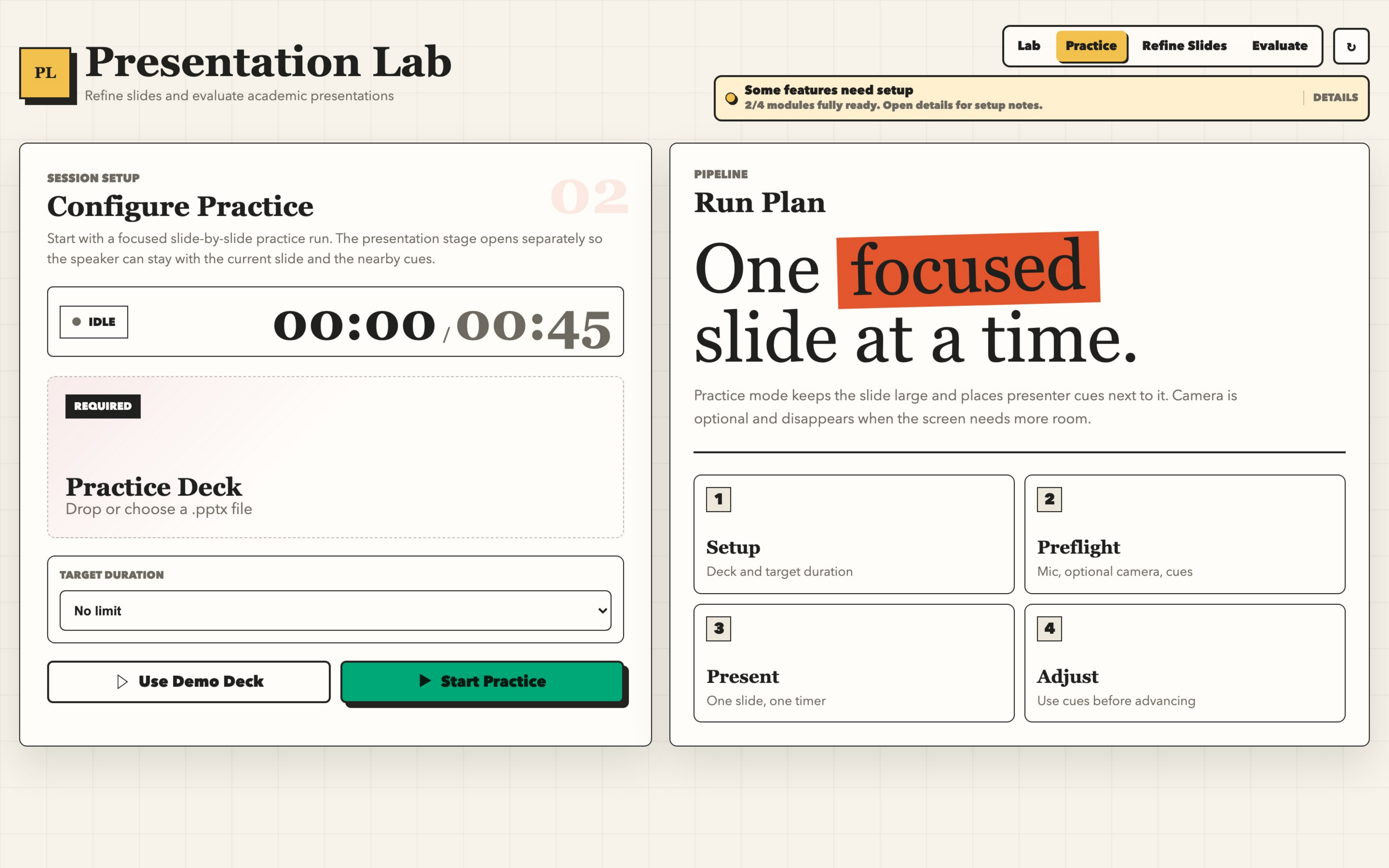}%
    }{%
      \fbox{\parbox[c][2.5cm][c]{0.95\linewidth}{\centering\textit{[Placeholder:\\demo2.pdf]}}}%
    }
    \parbox{\linewidth}{\centering(b) Practice Mode.}
  \end{minipage}\hfill
  \begin{minipage}[t]{0.32\textwidth}
    \centering
    \IfFileExists{demo3.pdf}{%
      \includegraphics[width=\linewidth,height=2.5cm,keepaspectratio]{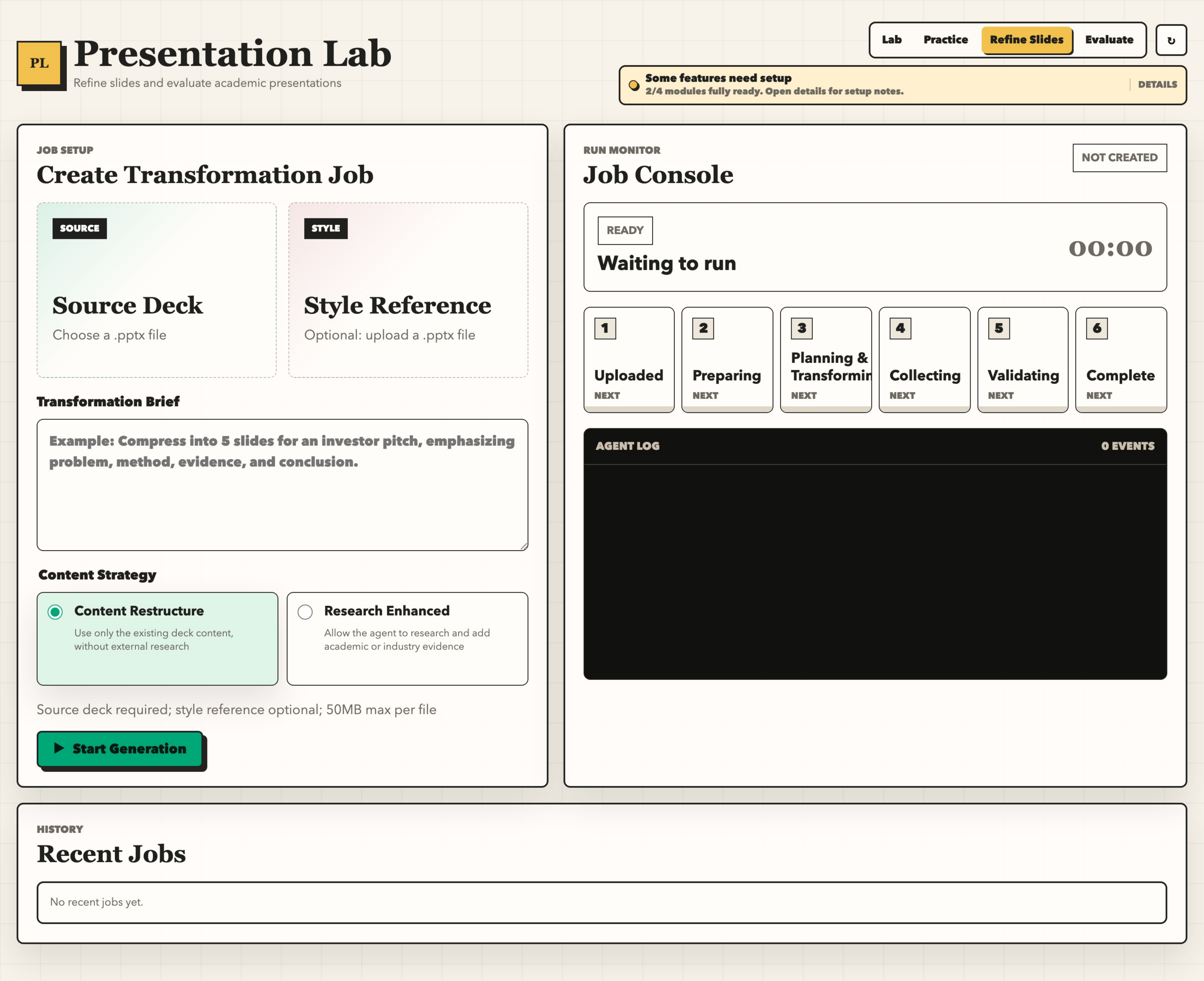}%
    }{%
      \fbox{\parbox[c][2.5cm][c]{0.95\linewidth}{\centering\textit{[Placeholder:\\demo3.pdf]}}}%
    }
    \parbox{\linewidth}{\centering(c) Slide deck refinement.}
  \end{minipage}
  \caption{Use-case walkthrough of PolyPresentation, from setup to slide deck refinement.}
  \label{fig:polypresentation-demo}
\end{figure}

Fig.~\ref{fig:polypresentation-demo} traces a typical session: the presenter selects a scenario and uploads a deck (a); a short Practice Mode pass surfaces speech rate, pauses, fillers, and keyword coverage (b); the slide deck refinement module then produces a refined deck and structured report (c).

\begin{figure}
  \centering
  \begin{minipage}[t]{0.48\textwidth}
    \centering
    \IfFileExists{demo4-a.pdf}{%
      \includegraphics[width=\linewidth,height=4.5cm,keepaspectratio]{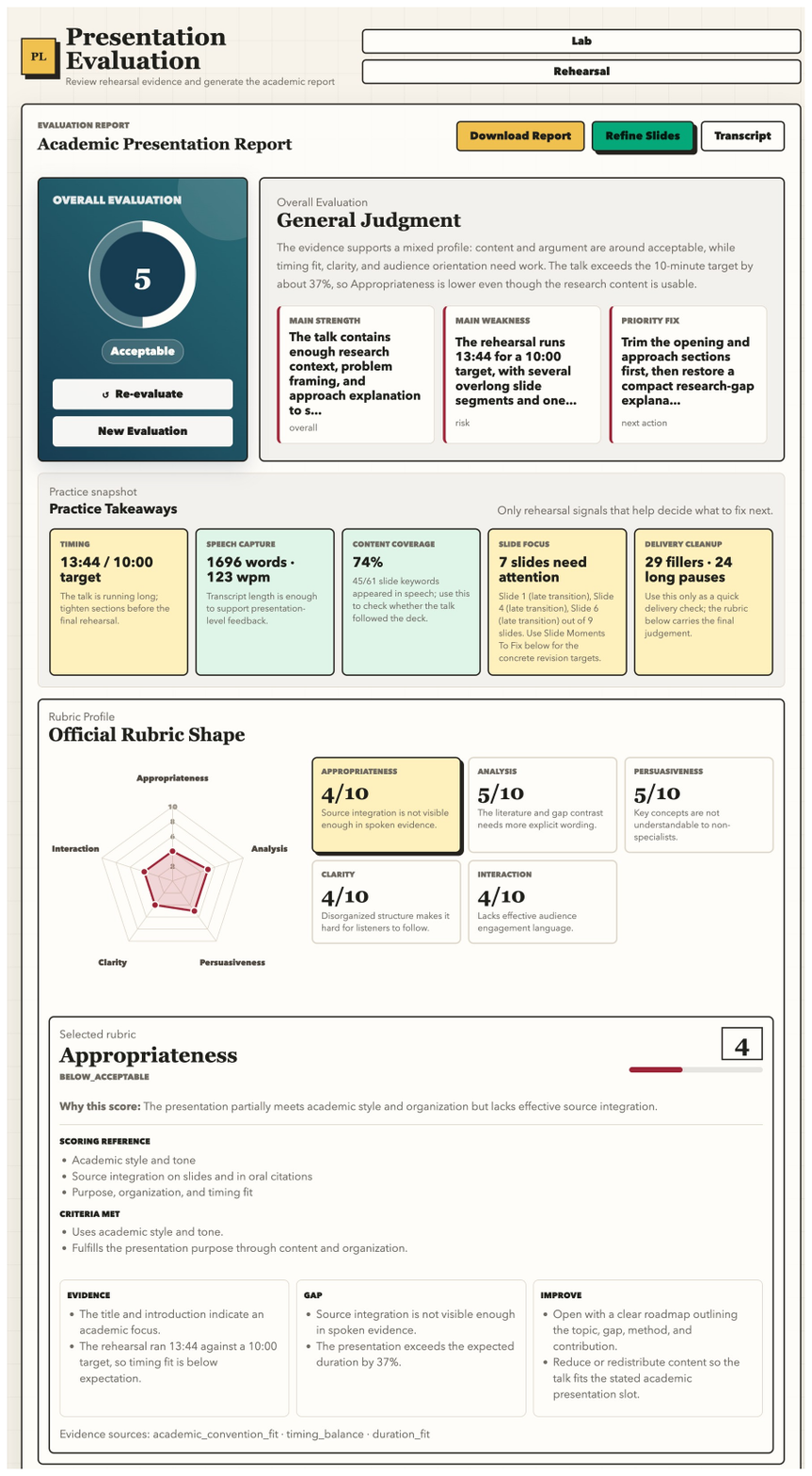}%
    }{%
      \fbox{\parbox[c][4.5cm][c]{0.95\linewidth}{\centering\textit{[Placeholder:\\demo4-a.pdf]}}}%
    }
    \parbox{\linewidth}{\centering(a) Report-level view: overall judgment, rubric scores, and rehearsed versus target timing.}
  \end{minipage}\hfill
  \begin{minipage}[t]{0.48\textwidth}
    \centering
    \IfFileExists{demo4-b.pdf}{%
      \includegraphics[width=\linewidth,height=4.5cm,keepaspectratio]{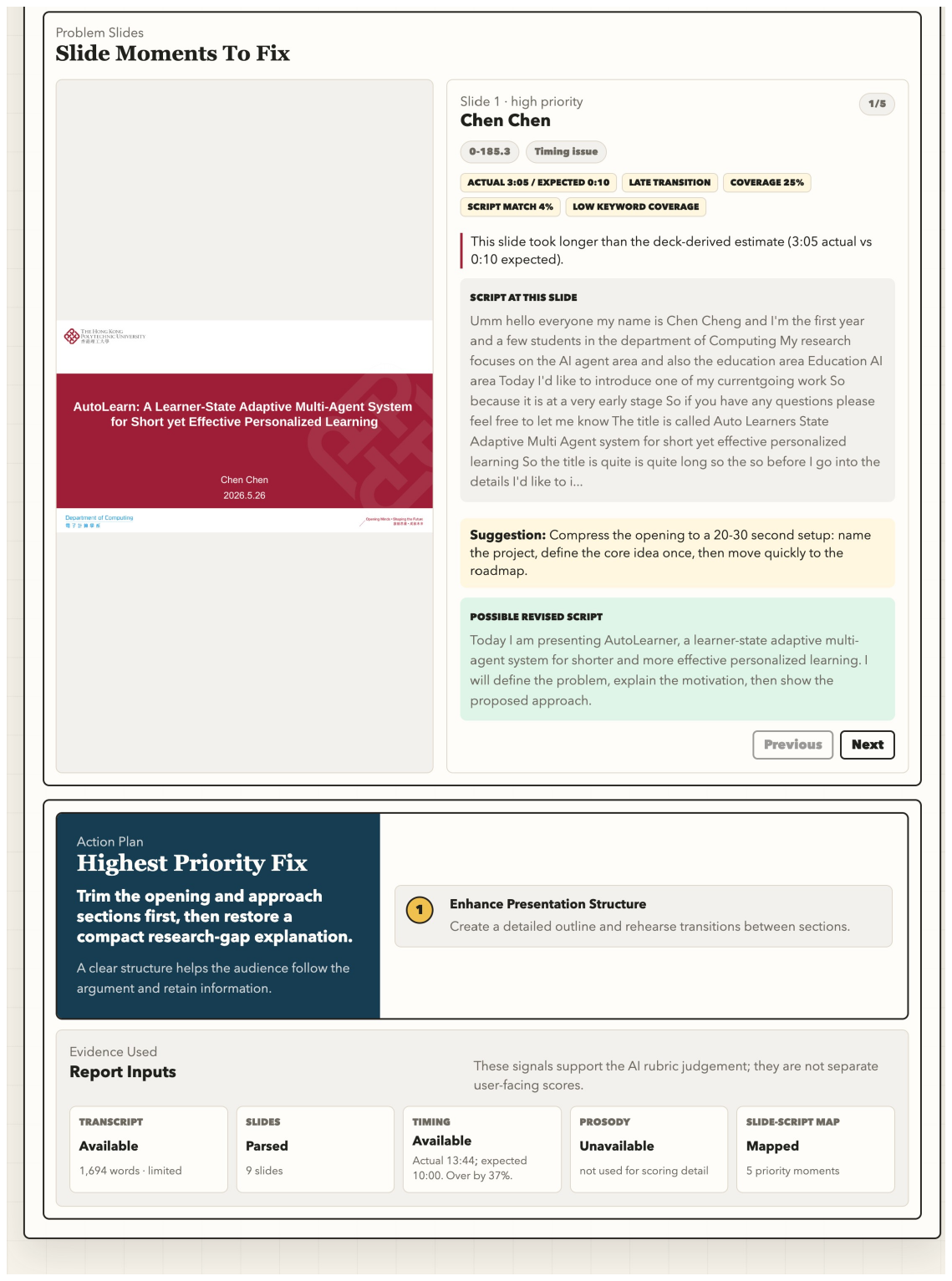}%
    }{%
      \fbox{\parbox[c][4.5cm][c]{0.95\linewidth}{\centering\textit{[Placeholder:\\demo4-b.pdf]}}}%
    }
    \parbox{\linewidth}{\centering(b) Detail view: per-slide diagnostics, Action Plan, and Report Inputs.}
  \end{minipage}
  \caption{The Evaluation Dashboard converts rubric findings into per-slide diagnostics and explicit next-round practice goals.}
  \label{fig:polypresentation-eval}
\end{figure}

Fig.~\ref{fig:polypresentation-eval} shows how the Evaluation Dashboard closes the loop. The report-level view (a) reports the overall judgment, per-rubric scores with a radar overlay, and rehearsed versus target duration. The detail view (b) lists per-slide diagnostics, suggested revisions, an Action Plan of next-round goals, and a Report Inputs panel marking which modalities were used or unavailable.

\section{Experiment and Evaluation}

We evaluate PolyPresentation on two axes: rubric-scoring alignment with a human reference (Sect.~\ref{sec:rubric-alignment}) and feedback quality against four baselines under a frozen comprehensive rubric scored by a GPT-5 judge (Sect.~\ref{sec:strict-rubric}).

\subsection{Rubric Alignment with Human Reference}\label{sec:rubric-alignment}

We assess rubric-scoring alignment on 20 academic-conference presentations using five dimensions on a 0--10 scale. \textit{Appropriateness} measures whether the presentation fits its audience, purpose, and constraints; \textit{Analysis} captures the depth and coherence of its reasoning; \textit{Persuasiveness} reflects how effectively claims are supported and communicated; \textit{Clarity} concerns the organization and comprehensibility of the presentation; and \textit{Interaction} assesses audience engagement and responses during Q\&A. We report Pearson $r$ for linear association, quadratic weighted kappa (QWK) for ordinal agreement, ICC(2,1) for absolute agreement, and mean absolute error (MAE) for score deviation.

\begin{table}
\caption{Agreement between PolyPresentation and human ratings across five criteria for 20 academic conference presentations (0--10 scale). Pooled statistics were calculated by combining the paired ratings across all five criteria. ``Hum'' and ``Plat'' denote mean human and platform scores, respectively. Boldface marks the highest $r$, QWK, and ICC(2,1), and the lowest MAE across criteria.}
\label{tab:rubric-alignment}
\centering
\small
\begin{tabular}{|l|c|c|c|c|c|c|}
\hline
\textbf{Dimension} & \textbf{Pearson $r$} & \textbf{QWK} & \textbf{ICC(2,1)} & \textbf{MAE} & \textbf{Hum} & \textbf{Plat} \\
\hline
Appropriateness & 0.860 & 0.857 & 0.863 & 0.25 & 7.00 & 6.95 \\
Analysis        & \textbf{0.910} & \textbf{0.903} & \textbf{0.907} & \textbf{0.20} & 6.30 & 6.40 \\
Persuasiveness  & 0.634 & 0.487 & 0.584 & 0.35 & 5.95 & 6.00 \\
Clarity         & 0.784 & 0.783 & 0.791 & 0.55 & 6.55 & 6.60 \\
Interaction     & 0.648 & 0.611 & 0.623 & 0.35 & 5.50 & 5.65 \\
\hline
\textbf{Pooled} & 0.836 & 0.830 & 0.831 & 0.34 & 6.26 & 6.32 \\
\hline
\end{tabular}
\end{table}

Table~\ref{tab:rubric-alignment} shows that PolyPresentation aligns well with the human ratings across the five presentation criteria. All three pooled agreement coefficients are at least 0.83, with a pooled MAE of 0.34, while the close pooled means (6.26 vs.\ 6.32) suggest limited overall scoring bias. Agreement is strongest for \textit{Analysis} and \textit{Appropriateness}, whereas \textit{Persuasiveness} and \textit{Interaction} show more moderate agreement. Overall, the results suggest that PolyPresentation agrees more closely with human ratings on evidence-based criteria than on more subjective aspects of presentation quality.

\subsection{Overall Feedback Quality}\label{sec:strict-rubric}

We compare PolyPresentation with four baselines: \textbf{PresentCoach} \cite{chen2025presentcoach}, \textbf{VLM} (Gemini-3.5 Flash), \textbf{LLM} (GPT-5, CHOP-inspired) \cite{cha2024chop}, and \textbf{Rule-based}. For each of the 20 samples, GPT-5 evaluated five anonymized feedback reports against the same multimodal evidence in three different orders. Scores were averaged across the three orders to reduce order bias. The rubric assesses seven dimensions: \textit{Validity}, how well the feedback claims are supported by evidence; \textit{Coverage}, the extent to which relevant issues are addressed; \textit{Impact}, the importance of the suggested improvements; \textit{Depth}, the quality of the reasoning behind the feedback; \textit{Actionability}, how specific and practical the recommendations are; \textit{Transfer}, how useful the feedback is for future rehearsals; and \textit{Organization}, how clearly the report is structured. The dimensions are weighted 20\%, 20\%, 15\%, 15\%,
15\%, 10\%, and 5\%.

\begin{table}[t]
\caption{Feedback-quality scores of five systems across seven dimensions, with a weighted overall score. Scores were averaged over 20 rehearsal samples using a fixed rubric evaluated by GPT-5 (1--10; higher is better). PresentCoach denotes our reproduction.}
\label{tab:rubric-overall}
\centering
\resizebox{\linewidth}{!}{%
\begin{tabular}{|l|c|c|c|c|c|c|c|c|}
\hline
\textbf{System} &
\textbf{Valid.} &
\textbf{Cover.} &
\textbf{Impact} &
\textbf{Depth} &
\textbf{Action.} &
\textbf{Transfer} &
\textbf{Org.} &
\textbf{Weighted Score} \\
\hline
PolyPresentation (ours)
& 6.97 & \textbf{8.15} & \textbf{7.68} & \textbf{7.50}
& \textbf{7.49} & \textbf{7.28} & \textbf{7.77} & \textbf{7.54} \\
PresentCoach (reprod.)
& \textbf{8.07} & 6.02 & 6.65 & 6.13
& 6.74 & 5.97 & 7.69 & 6.73 \\
VLM (Gemini-3.5 Flash)
& 5.41 & 6.38 & 6.41 & 5.83
& 6.22 & 5.69 & 6.81 & 6.04 \\
LLM (GPT-5, CHOP-inspired)
& 7.49 & 4.70 & 3.96 & 4.48
& 4.98 & 3.93 & 6.70 & 5.18 \\
Rule-based
& 4.90 & 4.18 & 4.24 & 3.81
& 3.94 & 3.97 & 6.31 & 4.33 \\
\hline
\end{tabular}}
\end{table}

Table~\ref{tab:rubric-overall} shows that PolyPresentation achieves the highest overall score (7.54), ranks first on 18 of 20 samples, and leads six of the seven dimensions. Its largest gains over PresentCoach occur in \textit{Coverage} (+2.13), \textit{Depth} (+1.37), and \textit{Transfer} (+1.31), indicating stronger support for comprehensive diagnosis, presentation revision, and subsequent rehearsal. PresentCoach leads \textit{Validity}, while VLM provides moderate multimodal feedback. The LLM baseline remains relatively strong in \textit{Validity} but offers less coverage and presentation-level guidance, and rule-based feedback performs lowest overall.

\section{Conclusion}
In this paper, we present \textbf{PolyPresentation}, a multimodal AI platform for slide-aware iterative presentation practice. The platform integrates slide-by-slide practice, full rehearsal, audience Q\&A, evidence-grounded feedback, and slide deck refinement into a unified practice loop. Experimental results suggest that PolyPresentation produces assessments broadly consistent with human judgment while translating rehearsal evidence into actionable guidance for subsequent practice.

This study nevertheless has several limitations. The evaluation was based on 20 academic-conference rehearsals and did not examine improvement over multiple practice rounds. Although rubric-scoring alignment was evaluated against human ratings, the feedback-quality comparison relied on GPT-5 as the judge. Because GPT-5 is also used in several PolyPresentation modules, this comparison may be subject to evaluation bias. Future work will include blinded human evaluation of feedback quality, multi-round user studies, and further investigation of system robustness, latency, and generalizability across presentation contexts.

\begin{credits}
\subsubsection{\ackname} This research work was conducted at the Research Institute for Artificial Intelligence of Things (RIAIoT) and The Institute for Higher Education Research and Development (IHERD) of PolyU.  This work was supported in part by the Hong Kong Research Grants Council Theme-based Research Scheme (No. T43-518/24-N), PolyU Internal Research Fund (No. BDZ3), and PolyU LTC Project (Grant TDLEG25-28/IICA/P/05, No. 48EM).
\end{credits}

\end{document}